# A Quantum Computational Learning Algorithm

Dan Ventura and Tony Martinez
Neural Networks and Machine Learning Laboratory (*http://axon.cs.byu.edu*)
Department of Computer Science
Brigham Young University
*dan@axon.cs.byu.edu*, *martinez@cs.byu.edu*

**Abstract** An interesting classical result due to Jackson allows polynomial-time learning of the function class DNF using membership queries. Since in most practical learning situations access to a membership oracle is unrealistic, this paper explores the possibility that quantum computation might allow a learning algorithm for DNF that relies only on example queries. A natural extension of Fourier-based learning into the quantum domain is presented. The algorithm requires only an example oracle, and it runs in $O(\sqrt{2^n})$ time, a result that appears to be classically impossible. The algorithm is unique among quantum algorithms in that it does not assume *a priori* knowledge of a function and does not operate on a superposition that includes all possible basis states.

## 1. Introduction

The field of computational learning theory (COLT) attempts to formalize the process of inductive learning and to develop bounds for classes of functions that are learnable in some formal sense. One particular function class that has received a great deal of attention is the class of binary functions that are representable in disjunctive normal form (DNF). This function class is very expressive and has resisted attempts to produce an algorithm for learning it under the rigorous PAC-learning model often used in computational learning theory. Recently, however, Jackson has produced an algorithm that learns DNF under the uniform distribution in polynomial time with access to a membership oracle (these terms will be defined shortly). This is an impressive theoretical result that relies on two key assumptions: uniform distribution of the function examples and access to a membership oracle. While the assumption of a uniform distribution is often reasonable in practice, access to a membership oracle usually is not.

The field of quantum computation (QC) investigates the power of the unique characteristics of quantum systems used as computational machines. Early quantum computational successes have been impressive, yielding results that are classically impossible [Sho97] [Gro96] [Hog96] [Ter98]. Thus, in an attempt to extend Jackson's result so that it does not rely on a membership oracle, this paper presents a Fourier-based quantum learning algorithm. Specifically, the algorithm is designed to find the large Fourier coefficients of boolean functions in polynomial time with access to only a classical example oracle. The algorithm is not completely successful because it fails to guarantee learning in polynomial time. However, it can be shown to learn in time $O(\sqrt{2^n})$. In contrast, classically estimating the Fourier coefficients requires $O(2^n)$ time. Therefore, although we have failed to accomplish the ultimate goal of polynomial-time learning using only an example oracle, we have nevertheless discovered a result that is both theoretically interesting (since it accomplishes something that seems classically impossible) and to some extent useful in a practical sense (since it extends the size of problems to which we can practically apply a Fourier-based learning scheme. We have, in essence, traded Jackson's impressive theoretical polynomial time bound for a practical assumption and a more modest time bound. It is also interesting to note that Grover's well-known quantum searching algorithm also provides the same $O(\sqrt{2^n})$ improvement over its classical equivalent [Gro96].

Section 2 provides a simplified overview of computational learning theory and briefly discusses work in COLT related to our results. Section 3 introduces quantum computation and some of its basic ideas and early successes. Since a familiarity with basic ideas from computational learning theory is assumed, only a few necessary remarks on the subject are provided here; quantum computation, on the other hand, is treated more carefully as it is likely that readers will be less familiar with it. Since neither of these subjects can be properly covered here, references for further study are also provided. Section 4 discusses Fourier-based learning methods in some detail. A Fourier-based, inductive quantum learning algorithm that only requires access to an example oracle is presented in section 5. Section 5 also discusses why the algorithm will not learn the function class DNF in polynomial time and then shows that it will learn in $O(\sqrt{2^n})$ time. The paper concludes with final remarks and directions for further research in section 6.

## 2. Computational Learning Theory

A rigorous approach to machine learning is traditionally traced back to Valiant [Val84], and the resulting computational learning theory has provided a formal basis for machine learning. In particular, the PAC (Probably Approximately Correct) model has yielded some nice theoretical results proving learnability of various function classes. Under this model an example oracle $E$ for a function $f$ may be queried for a random example of the form $x \to f(x)$ from $f$ according to a distribution $D$ that governs the frequency of the examples (basically this is equivalent to having access to a "large enough" training set). A function class $F$ is termed *strongly* PAC-learnable if there exists an algorithm $A$ such that for any $f \in F$ and for any $D$, $A$ produces a hypothesis $h$ such that $P(|h(x)-f(x)| > \varepsilon) < \delta$, and $A$ requires at most $m$ (where $m$ is polynomial in the size of the input) queries of $E$ and runs in time polynomial in $m$, $1/\delta$ and $1/\varepsilon$. Here $\varepsilon$ is called the error and $\delta$ the confidence. In other words, a function class is *strongly* PAC-learnable if there exists a learning algorithm that with high probability will produce a relatively accurate hypothesis in a polynomial amount of time. Another type of learning that may be defined is *weak* learning. A function class $F$ is *weakly* PAC-learnable if $F$ is PAC-learnable for $\varepsilon = 1/2 - 1/p(n,r)$, where $p$ is a fixed polynomial and $r$ is the size of $f$. That is, a function class is *weakly* PAC-learnable if there exists a learning algorithm that with high probability will produce a hypothesis that is at least slightly more accurate than random guessing.

However, the class DNF (disjunctive normal form) in which functions are expressed as a disjunction of clauses, each of which consists of a conjunction of literals (binary variables or their negation), has resisted efforts to develop provably efficient algorithms for its learnability. This is unfortunate because DNF is an extremely expressive class of functions, and it would therefore be significant to prove its learnability. Until recently, DNF as a general function class has resisted all attempts to find an algorithm that will learn it in the PAC sense; however, much work has been done regarding its learnability, and some encouraging results have been obtained for restricted subclasses of DNF [Aiz91] [Blu92] [Bsh93] [Kea87]. Currently, a particularly promising approach to the unrestricted class DNF is the use of discrete Fourier analysis in the development of machine learning algorithms, including work by Goldreich and Levin [Gol89], Kushilevitz and Mansour [Kus93], and Jackson [Jac94]. Jackson's work has yielded the most encouraging results so far concerning DNF, as he is able to produce the first positive learning results for the class of unrestricted DNF. Jackson's basic approach is to first note that binary functions may be represented as a discrete Fourier expansion and that all DNF functions may be at least weakly approximated using a single Fourier basis function whose coefficient is "large." His idea is to find those large coefficients by approximating the function using examples and to combine them in such a way as to produce a good hypothesis for the function. Jackson's learning algorithm, which he calls the Harmonic Sieve, combines techniques from discrete Fourier analysis due to Goldreich and Levin [Gol89] and Kushilevitz and Mansour [Kus93] with the boosting ideas of Shapire [Sha90] and Freund [Fre90] [Fre92]. The Harmonic Sieve guarantees strong learnability of DNF with two caveats: first, the function distribution $D$ must be uniform (actually this is relaxed somewhat, but it is still restrictive) and second, access to a membership oracle $M$ rather than to an example oracle $E$ is required. A membership oracle is like a black box version of the function to be learned, so that while an example oracle simply returns a random example upon being queried, a membership oracle can be queried for specific examples. In other words, when queried with $x$, the membership oracle must return $f(x)$. This characteristic makes a membership oracle strictly more powerful than an example oracle, and the Harmonic Sieve's dependence upon such an oracle, along with its restrictions on the distribution $D$, eliminates it from consideration as an algorithm for learning DNF in the PAC sense. Further, although Jackson's results are extremely impressive from a theoretical standpoint, they will likely yield very few practical results because, in general, access to a membership oracle is not realistic. On the other hand, access to an example oracle is much more realistic as this is basically equivalent to having access to a large training set.

Bshouty and Jackson [Bsh95] realized this and investigated using quantum computation to improve Jackson's work, just as is proposed here. The result was an extension to the Harmonic Sieve that depended upon a *quantum example* oracle rather than upon a *classical membership* oracle. Since the quantum example oracle is strictly less powerful than the classical membership oracle, this is a positive theoretical result. However, as the authors point out, it is unclear how to construct a quantum example oracle (other than perhaps by utilizing a classical membership oracle), and therefore their approach is again not useful in a practical sense. In contrast, the algorithm presented here goes one step further by depending only upon a *classical example* oracle, thus providing the possibility of practical algorithms for learning the class DNF.

## 3. Quantum Computation

Quantum computation is based upon physical principles from the theory of quantum mechanics (QM), which is in many ways counterintuitive. Yet it has provided us with perhaps the most accurate physical theory (in terms of predicting experimental results) ever devised by science. The theory is well-established and is covered in its basic form by many textbooks (see for example [Fey65]). Several necessary ideas that form the basis for the study of quantum computation are briefly reviewed here.

## 3.1 Linear Superposition

*Linear superposition* is closely related to the familiar mathematical principle of linear combination of vectors. Quantum systems are described by a wave function $\psi$ that exists in a Hilbert space [You88]. The Hilbert space has a set of states, $|\phi_i\rangle$, that form a basis, and the system is described by a quantum state $|\psi\rangle$,

$$|\psi\rangle = \sum_i c_i |\phi_i\rangle. \tag{1}$$

$|\psi\rangle$ is said to be in a linear superposition of the basis states $|\phi_i\rangle$, and in the general case, the coefficients $c_i$ may be complex. Use is made here of the Dirac bracket notation, where the ket $|\cdot\rangle$ is analogous to a column vector, and the bra $\langle\cdot|$ is analogous to the complex conjugate transpose of the ket. In quantum mechanics the Hilbert space and its basis have a physical interpretation, and this leads directly to perhaps the most counterintuitive aspect of the theory. The counter intuition is this -- at the microscopic or quantum level, the state of the system is described by the wave function $\psi$, that is, as a linear superposition of all basis states (i.e. in some sense the system is in all basis states at once). However, at the macroscopic or classical level the system can be in only a single basis state. For example, at the quantum level an electron can be in a superposition of many different energies; however, in the classical realm this obviously cannot be.

## 3.2 Coherence and decoherence

*Coherence* and *decoherence* are closely related to the idea of linear superposition. A quantum system is said to be coherent if it is in a linear superposition of its basis states. A result of quantum mechanics is that if a system that is in a linear superposition of states interacts in any way with its environment, the superposition is destroyed. This loss of coherence is called decoherence and is governed by the wave function $\psi$. The coefficients $c_i$ are called probability amplitudes, and $|c_i|^2$ gives the probability of $|\psi\rangle$ collapsing into state $|\phi_i\rangle$ if it decoheres. Note that the wave function $\psi$ describes a real physical system that must collapse to exactly one basis state. Therefore, the probabilities governed by the amplitudes $c_i$ must sum to unity. This necessary constraint is expressed as the unitarity condition

$$\sum_i |c_i|^2 = 1. \tag{2}$$

In the Dirac notation, the probability that a quantum state $|\psi\rangle$ will collapse into an eigenstate $|\phi_i\rangle$ is written $|\langle\phi_i|\psi\rangle|^2$ and is analogous to the dot product (projection) of two vectors. Consider, for example, a discrete physical variable called spin. The simplest spin system is a two-state system, called a spin-1/2 system, whose basis states are usually represented as $|\uparrow\rangle$ (spin up) and $|\downarrow\rangle$ (spin down). In this simple system the wave function $\psi$ is a distribution over two values (up and down) and a coherent state $|\psi\rangle$ is a linear superposition of $|\uparrow\rangle$ and $|\downarrow\rangle$. One such state might be

$$|\psi\rangle = \frac{2}{\sqrt{5}}|\uparrow\rangle + \frac{1}{\sqrt{5}}|\downarrow\rangle. \tag{3}$$

As long as the system maintains its quantum coherence it cannot be said to be either spin up or spin down. It is in some sense both at once. Classically, of course, it must be one or the other, and when this system decoheres the result is, for example, the $|\uparrow\rangle$ state with probability

$$|\langle\uparrow|\psi\rangle|^2 = \left(\frac{2}{\sqrt{5}}\right)^2 = .8. \tag{4}$$

A simple two-state quantum system, such as the spin-1/2 system just introduced, is used as the basic unit of quantum computation. Such a system is referred to as a quantum bit or *qubit*, and renaming the two states $|0\rangle$ and $|1\rangle$ it is easy to see why this is so.

### 3.3 Operators

*Operators* on a Hilbert space describe how one wave function is changed into another. Here they will be denoted by a capital letter with a hat, such as $\hat{A}$, and they may be represented as matrices acting on vectors. Using operators, an eigenvalue equation can be written $\hat{A}|\phi_i\rangle = a_i|\phi_i\rangle$, where $a_i$ is the eigenvalue. The solutions $|\phi_i\rangle$ to such an equation are called eigenstates and can be used to construct the basis of a Hilbert space as discussed in section 3.1. In the quantum formalism, all properties are represented as operators whose eigenstates are the basis for the Hilbert space associated with that property and whose eigenvalues are the quantum allowed values for that property. It is important to note that operators in quantum mechanics must be linear operators and further that they must be unitary so that $\hat{A}^\dagger\hat{A} = \hat{A}\hat{A}^\dagger = \hat{I}$, where $\hat{I}$ is the identity operator, and $\hat{A}^\dagger$ is the complex conjugate transpose, or adjoint, of $\hat{A}$.

### 3.4 Interference

*Interference* is a familiar wave phenomenon. Wave peaks that are in phase interfere constructively (magnify each other's amplitude) while those that are out of phase interfere destructively (decrease or eliminate each other's amplitude). This is a phenomenon common to all kinds of wave mechanics from water waves to optics. The well-known double slit experiment demonstrates empirically that at the quantum level interference also applies to the probability waves of quantum mechanics. As a simple example, suppose that the wave function described in (3) is represented in vector form as

$$|\psi\rangle = \frac{1}{\sqrt{5}}\begin{pmatrix}2\\1\end{pmatrix} \tag{5}$$

and suppose that it is operated upon by an operator $\hat{O}$ described by the following matrix,

$$\hat{O} = \frac{1}{\sqrt{2}}\begin{pmatrix}1 & 1\\1 & -1\end{pmatrix}. \tag{6}$$

The result is

$$\hat{O}|\psi\rangle = \frac{1}{\sqrt{2}}\begin{pmatrix}1 & 1\\1 & -1\end{pmatrix}\frac{1}{\sqrt{5}}\begin{pmatrix}2\\1\end{pmatrix} = \frac{1}{\sqrt{10}}\begin{pmatrix}3\\1\end{pmatrix} \tag{7}$$

and therefore now

$$|\psi\rangle = \frac{3}{\sqrt{10}}|\uparrow\rangle + \frac{1}{\sqrt{10}}|\downarrow\rangle. \tag{8}$$

Notice that the amplitude of the $|\uparrow\rangle$ state has increased while the amplitude of the $|\downarrow\rangle$ state has decreased. This is due to the wave function interfering with itself through the action of the operator -- the different parts of the wave function interfere constructively or destructively according to their relative phases just like any other kind of wave.

## 3.5 Entanglement

*Entanglement* is the potential for quantum states to exhibit correlations that cannot be accounted for classically. From a computational standpoint, entanglement seems intuitive enough -- it is simply the fact that correlations can exist between different qubits -- for example if one qubit is in the $|1\rangle$ state, another will be in the $|1\rangle$ state. However, from a physical standpoint, entanglement is little understood. The questions of what exactly it is and how it works are still not resolved. What makes it so powerful (and so little understood) is the fact that since quantum states exist as superpositions, these correlations somehow exist in superposition as well. When the superposition is destroyed, the proper correlation is somehow communicated between the qubits, and it is this "communication" that is the crux of entanglement. There are different degrees of entanglement and much work has been done on better understanding and quantifying it [Joz97] [Ved97]. It is interesting to note from a computational standpoint that quantum states that are superpositions of *only* basis states that are maximally far apart in terms of Hamming distance are those states with the greatest entanglement. For example, a superposition of only the states $|00\rangle$ and $|11\rangle$, which have a maximum Hamming spread, is maximally entangled. Finally, it should be mentioned that while interference is a quantum property that has a classical cousin, entanglement is a completely quantum phenomenon for which there is no classical analog.

## 3.6 Quantum Algorithms

The field of quantum computation, which applies ideas from quantum mechanics to the study of computation, was introduced in the mid 1980's [Fey86] [Ben82]. For a readable introduction to quantum computation see [Bar96]; for a more rigorous treatment see for example [Deu85]. The field is still in its infancy and very theoretical but offers exciting possibilities for the field of computer science -- the most important quantum algorithms discovered to date all perform tasks for which there are no classical equivalents. For example, Deutsch's algorithm [Deu92] is designed to solve the problem of identifying whether a binary function is constant (function values are either all 1 or all 0) or balanced (the function takes an equal number of 0 and 1 values). Deutsch's algorithm accomplishes the task in order $O(1)$ time, while classical methods require $O(2^n)$ time, where $n$ is the number of binary inputs to the function. Simon's algorithm [Sim97] is constructed to solve the following promise problem. A function $f:\{0,1\}^n \rightarrow \{0,1\}^n$ is guaranteed to be either a random 1-1 function or a periodic 2-1 function such that $f(x) = f(x \oplus s)$ for all $x$ for some $n$-bit string $s$ (where $\oplus$ denotes the bitwise exclusive OR). Here again an exponential speedup is achieved; however, admittedly, both these algorithms have been designed for artificial, somewhat contrived problems. Grover's algorithm [Gro96], on the other hand, provides a method for searching an unordered quantum database in time $O(\sqrt{2^n})$, compared to the classical lower bound of $O(2^n)$. Here is a real-world problem for which quantum computation provides performance that is classically impossible (though the speedup is less dramatic than exponential).

Finally, the most well-known and perhaps the most important quantum algorithm discovered so far is Shor's algorithm for prime factorization [Sho97]. This algorithm finds the prime factors of very large numbers in polynomial time, whereas the best known classical algorithms require exponential time. The implications for the field of cryptography are profound because many cryptographic systems, including the well-known RSA system [Riv78], depend upon the problem of prime factorization requiring exponential time. These quantum algorithms take advantage of the unique features of quantum systems to provide significant speedup over classical approaches.

It is worth mentioning that very recently several different groups have succeeded in physically realizing small-scale quantum computers and implementing some of the above mentioned algorithms with them [Jon98] [Chu98]. Also, work on quantum error correction has made impressive advances [Sho95] [Cor98] crucial to the construction of larger scale quantum computers. Therefore, although some formidable technological hurdles still exist, it is not unreasonable to suggest that quantum computational systems that perform nontrivial computation are much closer to realization than was thought possible even as recently as two or three years ago. In the mean time, it is important to develop a theory of quantum computation so that when the technology does become available, it may be exploited. Further, techniques and ideas that result from developing quantum algorithms may be useful in the development of new classical algorithms. Finally, the process of understanding and developing a theory of quantum computation provides insight and contributes to a furthering of our understanding and development of a general theory of computation.

## 4. Fourier-Based Learning

Here is given a brief description of Fourier-based learning as it applies to our approach. In what follows, the general Fourier-based learning approach used by Kushilevitz, Mansour, Jackson and others will for simplicity be referred to as KMJ. The subject of discrete Fourier analysis is well developed and is treated only very briefly here. For a more in depth presentation see any book on Fourier analysis, for example [Mor94].

### 4.1 Some ideas from Fourier analysis

A bipolar-valued binary function $f: \{0,1\}^n \to \{-1,1\}$ can be represented as a Fourier expansion (the expansion used here is actually a simplified Fourier expansion called a Walsh transform)

$$f(\vec{x}) = \sum_{\vec{a} \in \{0,1\}^n} \hat{f}(\vec{a}) \chi_{\vec{a}}(\vec{x}), \tag{9}$$

where the Fourier basis functions $\chi_{\vec{a}}(\vec{x})$ are defined as

$$\chi_{\vec{a}}(\vec{x}) = (-1)^{\vec{a}^{\mathrm{T}} \vec{x}} \tag{10}$$

and the Fourier coefficients being given by

$$\hat{f}(\vec{a}) = \frac{1}{2^n} \sum_{\vec{b} \in \{0,1\}^n} f(\vec{b}) \chi_{\vec{b}}(\vec{a}). \tag{11}$$

Actually, in the general case (11) should use $\chi^*$ (complex conjugate of $\chi$); however, in the simplified case considered here (bipolar rather than complex output), $\chi^* = \chi$. The KMJ method learns *f* in polynomial time by approximating a polynomial number of large Fourier coefficients in the context of a boosting algorithm due to Freund [Fre90] [Fre92]. In order to determine the set $\mathcal{A}$ of large coefficients, the method requires access to a membership oracle. The large coefficients, $\hat{f}(\vec{a})$ (for $\vec{a} \in \mathcal{A}$) are then approximated by

$$\tilde{\hat{f}}(\vec{a}) = \frac{1}{m} \sum_{\vec{x} \in \mathcal{T}} f(\vec{x}) \chi_{\vec{x}}(\vec{a}), \tag{12}$$

with $\mathcal{T}$ being a set of *m* carefully chosen examples of the form $\vec{x} \to f(\vec{x})$ with $\vec{x} \in \{0,1\}^n$ and $f(\vec{x}) \in \{-1,1\}$. Finally, using the fact that the function can be at least weakly approximated with a polynomial number of large coefficients (the set $\mathcal{A}$) and using (12) to approximate those coefficients, the function *f* may be approximated by

$$\tilde{f}(\vec{x}) = \sum_{\vec{a} \in \mathcal{A}} \tilde{\hat{f}}(\vec{a}) \chi_{\vec{a}}(\vec{x}). \tag{13}$$

We propose here a quantum algorithm that determines the set $\mathcal{A}$ using only an example oracle (i.e. a training set) rather than the membership oracle required by KMJ. In other words, the KMJ approach requires the ability to choose which examples will be used to learn the function (which in typical learning problems we can not do); on the other hand, the method that is proposed here makes no such requirement -- a standard training set suffices.

4.1.1 A Fourier example

A simple example will help illustrate the concept of a Fourier expansion. Let $n = 2$ and

$$f = \begin{cases} 00 \to 1 \\ 01 \to 1 \\ 10 \to -1 \\ 11 \to 1 \end{cases}.$$

To calculate the Fourier basis functions use (10) and for example,

$$\chi_{00}(00) = -1^{(00)\binom{0}{0}} = -1^0 = 1$$

and

$$\chi_{01}(11) = -1^{(01)\binom{1}{1}} = -1^1 = -1.$$

The other 14 values for the 4 Fourier functions are calculated similarly. Next, calculate the Fourier coefficients using (11). For example,

$$\hat{f}(11) = \frac{1}{4} \big( f(00)\chi_{00}(11) + f(01)\chi_{01}(11) + f(10)\chi_{10}(11) + f(11)\chi_{11}(11) \big).$$

That is,

$$\hat{f}(11) = \frac{1}{4} \big( (1)(1) + (1)(-1) + (-1)(-1) + (1)(1) \big) = \frac{1}{2},$$

and the other coefficients are found in the same manner. Finally, the Fourier expansion of *f* can be written using (9),

$$f(\vec{x}) = \hat{f}(00)\chi_{00}(\vec{x}) + \hat{f}(01)\chi_{01}(\vec{x}) + \hat{f}(10)\chi_{10}(\vec{x}) + \hat{f}(11)\chi_{11}(\vec{x}).$$

Which in this example evaluates to

$$f(\vec{x}) = \frac{1}{2}\chi_{00}(\vec{x}) - \frac{1}{2}\chi_{01}(\vec{x}) + \frac{1}{2}\chi_{10}(\vec{x}) + \frac{1}{2}\chi_{11}(\vec{x}),$$

and solving for any of the values 00, 01, 10, or 11 will result in the appropriate output for $f$. Now if instead of knowing $f$, we only have a training set such as

$$\mathcal{T} = \begin{cases} 00 \to 1 \\ 01 \to 1, \\ 10 \to -1 \end{cases}$$

then the coefficients are approximated using (12) instead of (11). For example now

$$\tilde{\hat{f}}(11) = \frac{1}{3}\left(f(00)\chi_{00}(11) + f(01)\chi_{01}(11) + f(10)\chi_{10}(11)\right),$$

which simplifies to

$$\tilde{\hat{f}}(11) = \frac{1}{3}\left((1)(1) + (1)(-1) + (-1)(-1)\right) = \frac{1}{3}.$$

The Fourier expansion is now approximated using (13) instead of (9).

$$\tilde{f}(\vec{x}) = \tilde{\hat{f}}(00)\chi_{00}(\vec{x}) + \tilde{\hat{f}}(01)\chi_{01}(\vec{x}) + \tilde{\hat{f}}(10)\chi_{10}(\vec{x}) + \tilde{\hat{f}}(11)\chi_{11}(\vec{x}),$$

which simplifies to

$$\tilde{f}(\vec{x}) = \frac{1}{3}\chi_{00}(\vec{x}) - \frac{1}{3}\chi_{01}(\vec{x}) + \frac{3}{3}\chi_{10}(\vec{x}) + \frac{1}{3}\chi_{11}(\vec{x}).$$

Now solving for 00, 01, or 10 will give 4/3, 4/3, and -4/3 respectively. While these are not the correct values, they are definitely the correct sign. On the other hand, solving for 11 results in 0, which is equivalent to "don't know".

## 4.2 Matrix formulation

The first step is to reformulate the Fourier equations in matrix form. Note that the $\chi_{\vec{a}}$ can be considered vectors in a $2^n$-dimensional space indexed by $\vec{x} \in \{0,1\}^n$. These $\chi_{\vec{a}}$ form an orthonormal basis for the function space with inner product

$$\langle \chi_{\vec{a}}, \chi_{\vec{b}} \rangle = E_x[\chi_{\vec{a}}(\vec{x})\chi_{\vec{b}}(\vec{x})] = \begin{cases} 1 & \text{if } \vec{a} = \vec{b} \\ 0 & \text{otherwise} \end{cases}. \tag{14}$$

Again, in the general case one of the $\chi$ in (14) should be $\chi^*$, but for the special case of bipolar outputs $\chi^* = \chi$. Now let **B** be the matrix formed by taking the $\chi_{\vec{a}}$ as the rows. Because of the orthonormality of the basis, the columns are also formed by the $\chi_{\vec{a}}$. Also, define $f$ as a vector in an $2^n$-dimensional space indexed by $\vec{x} \in \{0,1\}^n$ such that

$$f_{\vec{x}} = f(\vec{x}). \tag{15}$$

Then

$$\frac{1}{2^n}\mathbf{B}f = \hat{f} \tag{16}$$

gives the Fourier coefficients as a vector in a $2^n$-dimensional space indexed by $\vec{a} \in \{0,1\}^n$ so that

$$\hat{f}_{\vec{a}} = \hat{f}(\vec{a}). \tag{17}$$

To evaluate $\chi_{\vec{a}}(\vec{x})$, define $y$ as the $2^n$-vector indexed by $\vec{b} \in \{0,1\}^n$ such that

$$y_{\vec{b}} = \begin{cases} 1 & \text{if } \vec{b} = \vec{x} \\ 0 & \text{otherwise} \end{cases} \tag{18}$$

and calculate

$$\boldsymbol{B}y = \chi. \tag{19}$$

Now $\chi$ is a vector in a $2^n$-dimensional space indexed by $\vec{b} \in \{0,1\}^n$, the $\vec{a}$th element of which is equal to $\chi_{\vec{a}}(\vec{x})$. Finally, the Fourier representation of $f$ is given as

$$f(\vec{x}) = (\boldsymbol{B}y)^\dagger (\frac{1}{2^n} \boldsymbol{B}f). \tag{20}$$

Following the thinking of KMJ, $f$ may be approximated by approximating the Fourier coefficients using a set of $m$ examples, $\mathcal{T}$. Define $\tilde{f}$ as a $2^n$-vector indexed by $\vec{x} \in \{0,1\}^n$ such that

$$\tilde{f}_{\vec{x}} = \begin{cases} f(\vec{x}) & \text{if } \vec{x} \in \mathcal{T} \\ 0 & \text{otherwise} \end{cases}. \tag{21}$$

Then the Fourier representation of the approximate function is

$$\tilde{f}(\vec{x}) = (\boldsymbol{B}y)^\dagger (\frac{1}{m} \boldsymbol{B}\tilde{f}). \tag{22}$$

Now, using linear algebra gives

$$\tilde{f}(\vec{x}) = (\boldsymbol{B}y)^\dagger (\frac{1}{m} \boldsymbol{B}\tilde{f}) = \frac{1}{m} y^\dagger \boldsymbol{B}^\dagger \boldsymbol{B}\tilde{f} = \frac{2^n}{m} y^\dagger \tilde{f} = \begin{cases} \frac{2^n}{m} f(\vec{x}) & \text{if } \vec{x} \in \mathcal{T} \\ 0 & \text{otherwise} \end{cases}. \tag{23}$$

Recall that $f(\vec{x}) = \{-1,1\}$ and can never equal 0. Therefore, approximating $f$ by approximating all $2^n$ Fourier coefficients results in memorization of the training set (up to the sign of the output) and no generalization whatsoever on new inputs. As a consequence of this, it is clear that the generalization inherent in Fourier-based learning methods is due at least in part to the fact that not all the Fourier coefficients are used in the approximation.

### 4.2.1 A matrix version of the example

For clarity, the example of section 4.1.1 is repeated here using the matrix formulation. According to (15) the function $f$ is now written as a vector,

$$f_{\vec{x}} = \begin{pmatrix} 1 \\ 1 \\ -1 \\ 1 \end{pmatrix}.$$

The matrix $\boldsymbol{B}$ is

$$\boldsymbol{B} = \begin{pmatrix} 1 & 1 & 1 & 1 \\ 1 & -1 & 1 & -1 \\ 1 & 1 & -1 & -1 \\ 1 & -1 & -1 & 1 \end{pmatrix}$$

and using (16) the Fourier expansion of $f$ is now written

$$\frac{1}{4}Bf = \frac{1}{4}\begin{pmatrix} 1 & 1 & 1 & 1 \\ 1 & -1 & 1 & -1 \\ 1 & 1 & -1 & -1 \\ 1 & -1 & -1 & 1 \end{pmatrix}\begin{pmatrix} 1 \\ 1 \\ -1 \\ 1 \end{pmatrix} = \frac{1}{2}\begin{pmatrix} 1 \\ -1 \\ 1 \\ 1 \end{pmatrix}.$$

Now considering the case where the function is unknown but a training set is available, the approximate function represented by the training set is written as a vector using (21),

$$\tilde{f}_{\vec{x}} = \begin{pmatrix} 1 \\ 1 \\ -1 \\ 0 \end{pmatrix},$$

and the approximate Fourier transform is obtained from (22),

$$\frac{1}{3}B\tilde{f} = \frac{1}{3}\begin{pmatrix} 1 & 1 & 1 & 1 \\ 1 & -1 & 1 & -1 \\ 1 & 1 & -1 & -1 \\ 1 & -1 & -1 & 1 \end{pmatrix}\begin{pmatrix} 1 \\ 1 \\ -1 \\ 0 \end{pmatrix} = \frac{1}{3}\begin{pmatrix} 1 \\ -1 \\ 3 \\ 1 \end{pmatrix}.$$

Actually, (22) gives not just the approximate Fourier transform for $f$, but the method for calculating the individual values for the approximate expansion. For example, to calculate the approximate value for 11, use (18) to get

$$y = \begin{pmatrix} 0 \\ 0 \\ 0 \\ 1 \end{pmatrix}.$$

Then by (22)

$$f(\vec{x}) = \left(\begin{pmatrix} 1 & 1 & 1 & 1 \\ 1 & -1 & 1 & -1 \\ 1 & 1 & -1 & -1 \\ 1 & -1 & -1 & 1 \end{pmatrix}\begin{pmatrix} 0 \\ 0 \\ 0 \\ 1 \end{pmatrix}\right)^{\dagger} \frac{1}{3}\begin{pmatrix} 1 & 1 & 1 & 1 \\ 1 & -1 & 1 & -1 \\ 1 & 1 & -1 & -1 \\ 1 & -1 & -1 & 1 \end{pmatrix}\begin{pmatrix} 1 \\ 1 \\ -1 \\ 0 \end{pmatrix} = \begin{pmatrix} 1 & -1 & -1 & 1 \end{pmatrix}\frac{1}{3}\begin{pmatrix} 1 \\ -1 \\ 3 \\ 1 \end{pmatrix} = 0,$$

just as in example 4.1.1 and as proved in (23).

### 4.3 Quantum formulation

The next step is to extend the vector representation into the quantum domain. Given a set of $n$ (where $n$ is the length of the binary input $\vec{x}$ to $f$) qubits whose basis states, $|\vec{x}\rangle$, correspond to all the different values for $\vec{x}$, define

$$|f\rangle = \sum_{\vec{x}\in\{0,1\}^n} c_{\vec{x}}|\vec{x}\rangle, \tag{24}$$

where the amplitudes

$$c_{\vec{x}} = f(\vec{x}) = \{-1,1\} \tag{25}$$

are properly normalized according to (2), as a quantum state of the $n$ qubits that describes the function $f$. The domain of $f$ is encoded in the state labels and the range of the function in the phases of the amplitudes. Similarly, given a set of examples $\mathcal{T}$, define

$$\left|\tilde{f}\right\rangle = \sum_{\vec{x}\in\mathcal{T}} c_{\vec{x}} |\vec{x}\rangle, \tag{26}$$

where again the amplitudes are defined as in (25). This quantum state of the $n$ qubits describes the partial function $\tilde{f}$. Next, define the operator $\hat{B}$ as the matrix $\boldsymbol{B}$ and note that it acts on a quantum state, transforming it from the $|\vec{x}\rangle$ basis to the Fourier basis, $\chi_{\vec{a}}$. Now that transformation takes the form

$$\hat{B}|f\rangle = \left|\hat{f}\right\rangle, \tag{27}$$

or

$$\hat{B}\left|\tilde{f}\right\rangle = \left|\hat{\tilde{f}}\right\rangle, \tag{28}$$

slightly abusing the hat notation by using it to symbolize both an operator and the quantum state vector associated with the Fourier expansion of $f$. Similar to (19)

$$\hat{B}|\vec{x}\rangle = |\chi\rangle \tag{29}$$

results in a quantum state that contains in its amplitudes the values of all Fourier basis functions for the input $\vec{x}$. Finally,

$$f(\vec{x}) = \left\langle \hat{B}\vec{x} \middle| \hat{B}f \right\rangle \tag{30}$$

represents the full function and

$$\tilde{f}(\vec{x}) = \left\langle \hat{B}\vec{x} \middle| \hat{B}\tilde{f} \right\rangle \tag{31}$$

represents an approximation of the function from $\mathcal{T}$. Notice that we are still dealing with quantum amplitudes, and therefore the result of the calculation is not directly available to us. As it turns out, the quantum system will not be used to calculate $f(\vec{x})$ but only to determine the relative amplitudes of the Fourier coefficients -- those coefficients that are large will have "large" amplitudes.

4.3.1 A quantum example

Finally, the example of section 4.1.1 is repeated using the quantum formulation. This example is particularly interesting in that it emphasizes the difference of this quantum formulation from the previous two, indicating that we can carry the analogy only so far. According to (24) the function $f$ is now written as a quantum superposition,

$$|f\rangle = \frac{1}{2}|00\rangle + \frac{1}{2}|01\rangle - \frac{1}{2}|10\rangle + \frac{1}{2}|11\rangle.$$

The operator $\hat{B}$ is

$$\hat{B} = \frac{1}{2}\begin{pmatrix} 1 & 1 & 1 & 1 \\ 1 & -1 & 1 & -1 \\ 1 & 1 & -1 & -1 \\ 1 & -1 & -1 & 1 \end{pmatrix}$$

and using (27) the Fourier expansion of $f$ is now written

$$\hat{B}|f\rangle = \frac{1}{2}\begin{pmatrix} 1 & 1 & 1 & 1 \\ 1 & -1 & 1 & -1 \\ 1 & 1 & -1 & -1 \\ 1 & -1 & -1 & 1 \end{pmatrix}\frac{1}{2}\begin{pmatrix} 1 \\ 1 \\ -1 \\ 1 \end{pmatrix} = \frac{1}{2}|00\rangle - \frac{1}{2}|01\rangle + \frac{1}{2}|10\rangle + \frac{1}{2}|11\rangle.$$

Now considering the case where the function is unknown but a training set is available, the approximate function represented by the training set is written as a quantum superposition using (26),

$$|f\rangle = \frac{1}{\sqrt{3}}|00\rangle + \frac{1}{\sqrt{3}}|01\rangle - \frac{1}{\sqrt{3}}|10\rangle,$$

and the approximate Fourier transform is obtained from (28),

$$\hat{B}|\tilde{f}\rangle = \frac{1}{2}\begin{pmatrix} 1 & 1 & 1 & 1 \\ 1 & -1 & 1 & -1 \\ 1 & 1 & -1 & -1 \\ 1 & -1 & -1 & 1 \end{pmatrix}\frac{1}{\sqrt{3}}\begin{pmatrix} 1 \\ 1 \\ -1 \\ 0 \end{pmatrix} = \frac{1}{2\sqrt{3}}|00\rangle - \frac{1}{2\sqrt{3}}|01\rangle + \frac{3}{2\sqrt{3}}|10\rangle + \frac{1}{2\sqrt{3}}|11\rangle.$$

Note that the amplitudes are not the exact Fourier coefficients now, but are instead proportional to them. This is because of the unique statistical nature of quantum systems. It is also important to mention a subtle point here about using quantum computation to approximate the Fourier coefficients. Because of the normalization condition (2) we can not use the appropriate $1/m$ normalization of equations (12) and (22). The result of this is that equation (31) is not really *approximating* the Fourier coefficients of the original 2-valued function $f:\{0,1\}^n \to \{-1,1\}$ but rather is *computing* the Fourier coefficients of a 3-valued function $\tilde{f}:\{0,1\}^n \to \{-1,0,1\}$, which has the same "large" coefficients as $f$, except that they are normalized by an exponential factor. Fortunately, we will not need to use the quantum system to approximate the large coefficients but only to indicate which *are* the large coefficients. Still, this important difference will affect the efficiency of the algorithm, and we will address the problem it creates in section 5.1. Finally, using (30) and (31) gives us the ability to calculate function values and approximate function values, respectively. For example, to calculate the value for the binary string 11, use (30) to get

$$f(\vec{x}) = \langle \hat{B}\vec{x}|\hat{B}f \rangle = \frac{1}{2}(1 \quad -1 \quad -1 \quad 1)\frac{1}{2}\begin{pmatrix} 1 \\ -1 \\ 1 \\ 1 \end{pmatrix} = 1,$$

and use (31) to get

$$\tilde{f}(\vec{x}) = \langle \hat{B}\vec{x}|\hat{B}\tilde{f} \rangle = \frac{1}{2}(1 \quad -1 \quad -1 \quad 1)\frac{1}{2\sqrt{3}}\begin{pmatrix} 1 \\ -1 \\ 3 \\ 1 \end{pmatrix} = 0$$

just as in the previous examples and as proved in (23). We might also mention here that although mathematically these equations yield the correct values, from a physical standpoint, they are impossible to implement. Fortunately again, we do not need to be able to implement (30) or (31), but instead just need to be able to implement (28), which is physically realizable.

In the KMJ algorithm, the large coefficients are approximated using the instances in $\mathcal{T}$ as in (12) but recall that deciding which *are* the large coefficients first requires a membership oracle. The goal of the quantum formulation of the approximate Fourier expansion in (31) is to determine which are the large coefficients using only the training set $\mathcal{T}$ instead of using a membership oracle. The approach is to construct and observe the quantum system of equation (28). The intuitive idea is that since the large Fourier coefficients will be represented in this quantum state as amplitudes with "large" magnitude, they will have a "large" probability of being observed. Since observing the system will cause it to collapse to a single basis state corresponding to a single Fourier basis function, the process must be repeated and statistics kept for each basis state until a statistically significant measure of the relative amplitudes of the coefficients is achieved. How many times this observation process must be repeated to find a large coefficient will depend on how many large coefficients there are and on their relative magnitudes. This turns out to be the catch. As mentioned in 4.3.1, because of the nature of quantum systems, it is impossible to directly calculate equation (12). Instead, when using a quantum system the results of (12) are in essence multiplied by a normalization constant.

## 5. Finding a Large Coefficient by Quantum Fourier Sampling

Blum *et al*. have shown that under the uniform distribution, there exists a Fourier basis function that weakly approximates any DNF function $f$ [Blu94]. In other words, there exists a Fourier basis function, $\hat{f}_W$, that agrees with $f$ on at least 1-$\varepsilon$ inputs, where

$$\varepsilon = \frac{1}{2} - \frac{1}{p(n,s)} \qquad (32)$$

with $p$ being some fixed polynomial, $n$ being the number of inputs to $f$, and $s$ being the size of the representation of $f$. This implies that the magnitude of the Fourier coefficient for the weak approximator is

$$\left|\hat{f}_W\right| = O(\frac{1}{p(n,s)}), \qquad (33)$$

and using Chernoff bounds [Hag89] this can be closely approximated using a polynomial number $m$ of examples drawn randomly from an example oracle, if we know which of the $2^n$ basis functions it is. In other words, since this coefficient is relatively large, if we had some way of sampling the Fourier coefficient distribution, it should be readily identifiable with a polynomial number of samples. Following this line of thinking, the results developed in section 4 suggest a simple algorithm for finding this large Fourier coefficient which is presented in figure 1.

> Repeat until confidence $> \delta$
>     form $|\tilde{f}\rangle$, a quantum superposition representing the training set
>     compute the Fourier transform on the superposition using the operator $\hat{B}$
>     observe the system to produce a random sample of the coefficient distribution
> Approximate the most commonly observed coefficient classically

**Figure 1. Algorithm for finding large Fourier coefficients**

There are four ingredients for the algorithm, the first three of which are performed on a quantum computer while the fourth is performed on a classical computer . In order for the algorithm to run in polynomial time, all of them must be shown to be polynomial. Specifically, we must address: 1) The construction of the state $|\tilde{f}\rangle$; 2) The implementation of the operator $\hat{B}$; 3) The number of times this process must be repeated in order to identify a large coeffient; and 4) The classical approximation of that coefficient. Constructing the state $|\tilde{f}\rangle$ is nontrivial and a method for doing so in $O(mn)$ time is detailed in [Ven98]. Implementing $\hat{B}$ turns out to be extremely easy on a quantum computer, and it is in fact the basis of most quantum algorithms discovered to date. Computing the Walsh transform of a quantum state is accomplished simply by applying the elementary quantum operator

$$\hat{H} = \frac{1}{\sqrt{2}}\begin{bmatrix} 1 & 1 \\ 1 & -1 \end{bmatrix} \tag{34}$$

to each qubit in parallel (see for example [Gro96]). In other words, $\hat{B} = \hat{H} \otimes \cdots \otimes \hat{H}$, where $\otimes$ is the tensor product (direct matrix product) and $\hat{H}$ appears in the product $n$ times, one for each input. Approximating a coefficient is easy on a classical computer and can be accomplished using equation (12) and a polynomial number of function examples.

## 5.1 A Quantum Quirk in Estimating Coefficients

This leaves the question of how many times we must repeat the Fourier sampling process in order to find, with high probability, a large coefficient, and in fact, this is a difficult point. To see why, consider the following. The unitarity condition of equation (2) requires that the sum of the squares of the coefficients in a quantum state be equal to 1. For convenience we reproduce the equation here, recalling that the *quantum* coefficients $c_i$ are within a multiplicative constant of being equal to the *Fourier* coefficients $\hat{f}(\vec{a})$.

$$\sum_i |c_i|^2 = 1. \tag{35}$$

As mentioned in 4.3.1, what the Fourier sampling algorithm is really doing is not approximating the Fourier coefficients of $f:\{0,1\}^n \to \{-1,1\}$ but rather is calculating the Fourier coefficients of $\tilde{f}:\{0,1\}^n \to \{-1,0,1\}$. Another way to look at this is to say that the Fourier sampling algorithm *is* approximating the coefficients of $f:\{0,1\}^n \to \{-1,1\}$ but with a normalization constant of $1/\sqrt{2^n}$ rather than the $1/m$ required by equation (12). (Either viewpoint will produce the same result.)

Because of this, in contrast to the nice polynomial sized Fourier coefficient given in (33), the size of the corresponding quantum coefficient $c_W$, is

$$c_W = O\left(\frac{\sqrt{m/2^n}}{p(n,s)}\right) = O\left(\sqrt{\frac{m}{2^n}}\right), \tag{36}$$

and its square is thus

$$(c_W)^2 = O\left(\frac{m/2^n}{p^2(n,s)}\right) = O\left(\frac{m}{2^n}\right). \tag{37}$$

Therefore equations (35-37) require that the number $C$ of nonzero coefficients be at least

$$1 = C\left[O\left(\sqrt{\frac{m}{2^n}}\right)\right]^2 \Rightarrow C = O\left(\frac{2^n}{m}\right). \tag{38}$$

Also, the smallest possible quantum coefficient, $c_S$ is

$$c_S = \frac{1}{\sqrt{2^n}}. \tag{39}$$

Therefore, if the number of examples $m$ is significantly less than $2^n$, which will be the case if we require a polynomial-time algorithm, then there are an exponential number of nonzero coefficients in the quantum state with the largest only a factor of $\sqrt{m}$ larger than the smallest. (When the system is observed, this difference is magnified to a factor of $m$; however this is not significant in the end result.) This means that $O(2^n)$ samples of the amplitude distribution (obtained by observing the quantum system) will be required to discriminate the "large" coefficients from the "small" ones.

Another way to look at this is to say that the vector representing the function is very sparse. As a result, it correlates well with almost every Fourier basis function and therefore, the "large" coefficients are not very much larger than the "small" ones. In fact, as mentioned above, this is unavoidable in a quantum system because of the requirement that all operators be unitary. Therefore, the quantum system is attempting to calculate information about an exponential number of Fourier coefficients using only a polynomial number of function examples. Because of the unitarity requirement this polynomial amount of information gets normalized by an exponential instead of the polynomial necessary for closely approximating the coefficients as in (12). Because of the normalization inherent in unitary quantum processes, it is unlikely that a quantum algorithm can be developed to overcome this problem. However, this is certainly not a proof that it is impossible.

There is an equally valid classical analysis based upon Parseval's identity that will yield the same result, though some of the intermediate results are different by $O(\sqrt{\cdot})$.

## 5.2 A NonClassical Result -- Finding Large Coefficients in $O(\sqrt{2^n})$ Time

If the quantum algorithm of figure 1 requires exponential time to identify the large quantum coefficients, it is certainly no better than a classical algorithm that simply approximates each Fourier coefficient using equation (12) and thus also requires $O(2^n)$ time. However, the algorithm actually requires only $O(\sqrt{2^n})$ steps in order to identify the large quantum coefficients and thus still represents a significant improvement over the classical approach. To see why, choose the number of examples $m$ drawn from an example oracle to be

$$m = \sqrt{2^n}. \tag{40}$$

Then by equation (36) the quantum coefficient $c_W$ which corresponds to the Fourier function that weakly approximates $f$ is

$$c_W = O\left(\frac{\sqrt{m/2^n}}{p(n,s)}\right) = O\left(\frac{\sqrt{\sqrt{2^n}/2^n}}{p(n,s)}\right) = O\left(\frac{1}{2^{n/4}}\right), \tag{41}$$

and using equation (37) its square is

$$(c_W)^2 = O\left(\frac{m/2^n}{p^2(n,s)}\right) = O\left(\frac{\sqrt{2^n}/2^n}{p^2(n,s)}\right) = O\left(\frac{1}{\sqrt{2^n}}\right). \tag{42}$$

Using equation (38) the number $C$ of nonzero coefficients is now at least

$$1 = C\left[O\left(\frac{1}{2^{n/4}}\right)\right]^2 \Rightarrow C = O\left(\frac{1}{\sqrt{2^n}}\right). \tag{43}$$

Since the smallest possible quantum coefficient $c_S$ is still give by equation (39), we now have a situation in which there are still an exponential number of nonzero coefficients in the quantum state, but now the largest is a factor of $2^{n/4}$ larger than the smallest. When the system is observed, this difference is magnified to a factor of $\sqrt{2^n}$, and therefore $O(\sqrt{2^n})$ samples of the amplitude distribution will suffice to identify the large coefficient.

## 5.3 Analysis and discussion

Using some concrete numbers, assume that $n = 30$ and that therefore $m = \sqrt{2^{30}} = 2^{15} = 32768$, numbers which can represent a very non-trivial learning task. Then the algorithm requires $O(2^{15}) < 10^5$ operations. For comparison, [Bar96] gives estimates of how many operations might be performed before decoherence for various possible physical implementation technologies for the qubit. These estimates range from as low as $10^3$ (electron GaAs and electron quantum dots) to as high as $10^{13}$ (trapped ions), so our example falls comfortably into this range, even near the low end of it. Further, the algorithm would require only 61 qubits. (Although it appears that the algorithm presented here requires only $n$ qubits, the algorithm depends on a method for representing the training set as a quantum state. As mentioned before, an explicit algorithm for constructing such a quantum state exists [Ven98]; however it requires $2n+1$ qubits.)

In contrast, Shor's algorithm requires hundreds or thousands of qubits to perform an interesting factorization. For example, [Ved96] gives estimates for the number of qubits needed for modular exponentiation, which dominates Shor's algorithm -- anywhere from $7n+1$ down to $4n+3$. For a 512 bit number (which RSA actually claims may not be large enough to be safe anymore, even classically), this translates into anywhere from 3585 down to 2051 qubits. As for elementary operations, they claim $O(n^3)$, which would be in this case $O(10^8)$. Therefore, the algorithm presented here requires orders of magnitude fewer operations and qubits than Shor's in order to perform significant computational tasks. This is an important result since quantum computational technology is still immature, and maintaining and manipulating the coherent superposition of a quantum system of 60 qubits should be attainable sooner than doing so for a system of 2000 qubits. In this sense, the algorithm compares nicely with other quantum algorithms.

As mentioned earlier, the algorithm also compares favorably with classical approaches to finding large coefficients. Jackson's Harmonic Sieve runs in polynomial time; however, as discussed in section 2 this algorithm requires a membership oracle, or black box access to the function *f* that we are trying to learn, in order to find the large coefficients. In cases where this is not realistic, the algorithm presented here provides an alternative that requires only an example oracle. There are no known classical algorithms for finding large coefficients that use only an example oracle and run any faster than $O(2^n)$, so again the algorithm presented here represents a significant improvement.

## 6. Concluding Comments

This paper presents a quantum computational learning algorithm that takes advantage of the unique capabilities of quantum computation to produce a significant result in both the field of computational learning and the field of quantum computation. The main result of the paper is a quantum computational learning algorithm for learning the function class DNF over the uniform distribution using only an example oracle in time significantly faster than any known classical algorithm operating under the same constraints. In other words, the paper makes an important contribution to both the field of computational learning theory and to the field of quantum computation -- producing both a new learning theoretic result and a new quantum algorithm that accomplishes something that no classical algorithm has been able to do. Further, the algorithm is unique among quantum computational algorithms as it does not assume *a priori* knowledge of a function *f* and does not operate on a quantum supersposition that includes all possible basis states.

Further, the paper introduces a promising new field to which quantum computation may be applied to advantage -- that of computational learning. In fact, it is the authors' opinion that this application of quantum computation will, in general, demonstrate much greater returns than its application to more traditional computational tasks (though Shor's algorithm is an obvious exception). We make this conjecture because results in both quantum computation and computational learning are by nature probabilistic and inexact, whereas most traditional computational tasks require precise and deterministic outcomes.

One obvious area for future work is, of course, the physical implementation of the algorithm in a real quantum system. As mentioned in sections 3.6 and 5.3, the fact that this algorithm requires very few qubits for non-trivial problems combined with the recent advances in quantum technology suggests that the realization of quantum computers performing useful computation may be possible in the near future. A second important topic for future work is to determine whether or not the improvement obtained here over classical methods is optimal. If it is not, future work may produce further gains. In the mean time, a simulation of the quantum algorithm is being developed to run on a classical computer at the cost of an exponential slowdown in the size of the learning problem. Thus, learning problems that are non-trivial and yet small in size will provide interesting study in simulation. Various methods of using and combining the large Fourier coefficients are currently being investigated. One method, of course, is to use the boosting method that Jackson uses in his Harmonic Sieve, and we are considering other methods of combination as well. We are also investigating using higher spin systems for learning nonbinary problems. Another obvious and important area for future research is investigating further the application of quantum computational ideas to the fields of computational learning theory and machine learning -- the discovery of other quantum computational learning algorithms.

As a final comment, Grover's well-known algorithm for quantum search also provides a $O(\sqrt{2^n})$ improvement over classical approaches to the same problem. Grover's result is particularly important because in the case of searching an unordered database, the $O(2^n)$ bound for classical approaches is tight -- there is no classical way to do better. In the case of learning DNF using only an example oracle, it is still an open problem whether or not a classical algorithm exists that is better than $O(2^n)$; however, it is generally suspected that no such algorithm exists. This provides fuel for speculation that perhaps there exists a quantum information theoretic law that bounds the improvement of quantum algorithms over classical algorithms by $O(\sqrt{2^n})$. According to Shor's paper, the best classical algorithm for prime factorization of an integer that can be represented in $n$ bits runs in time $O(g)$, where $g = e^{n^{1/3}(\log n)^{2/3}}$ [Len93]. Since $g = O(\sqrt{2^n})$ but $g \neq \Theta(\sqrt{2^n})$, Shor's algorithm in some sense does not achieve as much of a speedup as do Grover's algorithm and the algorithm presented here.